\documentclass[]{article}
\usepackage{amsmath}
\usepackage{amssymb}
\newcommand{\be}{\begin{equation}}
 \newcommand{\ee}{\end{equation}}

\newtheorem{theorem}{Theorem}[section]
 
\newtheorem{lemma}[theorem]{Lemma}

 \newtheorem{defi}[theorem]{Definition}
\newtheorem{definition and theorem}[theorem]{Definition and Theorem}
\newtheorem{remark}[theorem]{Remark}
 \newtheorem{*remark}[theorem]{$^* $Remark}

\newtheorem{*exercise}[theorem]{$^* $Exercise}
\newtheorem{**exercise}[theorem]{$^{** } $Exercise}

\begin{document}
\title{Value-at-Risk and Expected Shortfall for Quadratic Portfolio of Securities with Mixture of Elliptic Distributed Risk Factors  }

\author{Jules SADEFO KAMDEM
\thanks{ Paper based on the M. Jules SADEFO KAMDEM Universit\'e de Reims Phd
Thesis.\quad Author address:  BP 1039 Moulin de la Housse 51687
Reims Cedex FRANCE. sadefo@univ-reims.fr  }
\thanks{The author is currently  temporary lecturer to the Mathematics Department of the ( Universit\'e d'evry val d'essonne FRANCE).}
\\
 Laboratoire de Math\'ematiques\\
 CNRS UMR 6056\\
 Universit\'e De Reims \\
 }

 \maketitle
\begin{abstract}
        Generally, in the financial literature, the notion of  quadratic VaR is implicitly  confused with
        the Delta-Gamma VaR, because more authors dealt with
 portfolios that contained derivatives instruments. In this paper,
we postpone to estimate both the expected shortfall and
Value-at-Risk of a quadratic portfolio of securities (i.e
equities) without the Delta and Gamma Greeks, when the
 joint  log-returns changes with multivariate elliptic distribution.  To illustrate
our method, we give special attention to mixture of normal
distributions, and mixture of
 Student t-distributions.
\end{abstract}

\section{\bf Introduction}
\medskip

           Value-at-Risk is a market risk management tool that permit  to measure the maximum
 loss of the portfolio   with certain confidence probability $1-\alpha$, over a certain time horizon such as one day. Formally ,
            if the price of portfolio's $ P(t, S(t))$ at time t is a random variable where $S(t)$ represents a vector
            of risk factors at time t, then VaR be  implicitly given by the formula
$$ Prob\{ -P(t,S(t))+P(0,S(0)) > VaR_\alpha \}=\alpha . $$

  Generally, to estimate the $VaR_{\alpha}$ for portfolios depending non-linearly on the return
  , or portfolios of non-normally distributed assets, one turns to Monte Carlo
  methods. Monte Carlo methodology has the obvious advantage of being almost universally
  applicable, but has the disadvantage of being much slower than
  comparable parametric methods, when the latter are available.

     In this paper we are concerned with the numerical
estimation of the losses that the portfolio of equities faces due
to the market, as a function of the future values of S. Following
the quadratic Delta-Gamma Portfolio, we introduce the notion of
quadratic portfolio of equities due to the analytic approximation
of Taylor in $2^{nd}$ order of log-returns for very small
variations of time. Quadratic approximations  have also been the
subject of a number of papers dedicated to numerical computations
for VaR ( but these have been done for portfolio that contains
derivatives instruments). We refer the reader to Cardenas and
al.(1997) \cite{CFKM} for a numerical method to compute quadratic
VaR using fast Fourier transform . Note that in \cite{BCQS},
Brummelhuis, Cordoba, Quintanilla and Seco have dealt with the
similar problem, but their work have been done analytically for
Delta-Gamma Portfolio VaR when the joint underlying risk factors
 follow a normal distribution which is a particular case of elliptic distribution .   All our
calculus will be done according to  the assumption that  the joint
securities (i.e equities) log-returns follow an elliptic
distribution. To illustrate our method, we will take some examples
of elliptic distributions as mixture of {\em{ multivariate
$t$-student}} and mixture of normal distributions. Note also that,
Following RiskMetrics,  Sadefo-Kamdem \cite{S2}(2003) have
generalized the notion of $\Delta$-normal VaR by introducing the
notion of $\Delta$-Elliptic VaR ,  with special attention to
$\Delta$-Student VaR , but this concerned the linear portfolio. In
this paper, we  will do the same for nonlinear
quadratic portfolios without derivatives instruments.\\

   The rest of the paper is organized as follows: In section
2, we introduced the notion of quadratic portfolio of
securities(i.e equities) due to the $2^{nd}$ order Taylor
approximation of log returns . Our calculus is made with the more
generalized assumptions that the joint underlying log-returns
follow an elliptic distribution. That is why in section 3,
following \cite{EMS}, we recall the definition of elliptic
distribution and we show that under the hypothesis of elliptic
distribution the VaR estimation of such portfolios is reduced to a
multiple integral equation. Next following  the paper of Alan Genz
\cite{G} , we recall the notions of symmetric interpolar rules for
multiple integrals over hypersphere and we use this method to
reduce our problem to one dimensional integral equation. In the
same section, we illustrate our method by giving an explicit
equation with solution VaR ( Value-at-Risk ) when the joint
log-returns follow some particular mixture of elliptic
distributions named mixture of multivariate Student
$t$-distributions or the mixture of normal distributions, in these
cases the VaR estimation is reduced to finding the zero's of a
certain specials functions. In section 5 we treats the expected
shortfall for general elliptic quadratic portfolios of securities
without derivatives instruments and we illustrate with the special
case of normal distribution. Finally, in section 6 we give a
 conclusion.

\section{Quadratic Portfolio of Securities(i.e Equities)}

   A portfolio of n securities is a vector $\theta \in
\mathbb{R}^{n}$; the component $\theta_{i}$ represents  the number
of holdings of the $i^{th}$ instruments, which in practise does
not need to be an integer.
  So at time t the price of the portfolio of n securities is given by:
\be P(t)=\sum_{i=1}^{n} \theta_{i}S_{i}(t)  \label{e1}  \ee where
$ S(t)=( S_{1}(t) ,\ldots, S_{n}(t)) $ such that
$$
P(t) - P(0) =\sum_{i=1}^{n} \theta_i   ( S_{i}(t) -S_{i}(0) ) =
\sum_{i=1}^{n}S_{i}(0) . \theta_{i} . (\frac{S_{i}(t)}{S_{i}(0)}
-1 ) \label{e3} $$ For small fluctuations of time and market, we
assume that log-return is given by :

\be log(S_{i}(t)/ S_{i}(0)) = \eta_{i}(t) \label{eq2} \ee
 therefore
$$S_{i}(t) - S_{i}(0) = S_{i}(0) ( \frac{S_{i}(t)}{S_{i}(0)} - 1)
=  S_{i}(0) (exp(\eta_{i}(t)) - 1)  $$
 Then we have that
$$S(t)=(S_{1}(0) exp(\eta_{1}), \ldots,S_{n}(0) exp(\eta_{n})). $$
 By using  Taylor's expansion of the exponential's due to  small
fluctuations of returns with a time, we have that :
\be
exp(\eta_{i}(t)) - 1 \approx  \eta_{i}(t) +
\frac{{\eta_{i}(t)}^{2}}{2}. \label{eq3}
\ee
 If we assume that  $\eta =(\eta_{1},\ldots,\eta_{n})$ is an elliptic distribution note by $ N_{n}(\mu
,\Sigma ,\phi )$ then

\be P(t) -P(0) =\sum_{i=1}^{n}S_{i}(0)\cdot \theta_{i}\cdot
(exp(\eta_{i}(t)) - 1  )  \approx  \sum_{i=1}^{n} S_{i}(0)
.\theta_{i} ( \eta_{i}(t) + \frac{{\eta_{i}(t)}^{2}}{2} ).
\label{e6} \ee

By  following the usual convention of recording portfolio losses
by negative numbers, but stating the Value-at-Risk as a  positive
quantity of money, The $VaR_{\alpha}$ at confidence level of
$1-\alpha$ is given by solution of the following equation:
$$Prob\{ |P(t) - P(0) | \geq VaR_{\alpha} \} =\alpha $$
In the probability space of losses $|P(t) - P(0) | =-P(t) + P(0)
$, therefore recalling (\ref{e6}), we have

 \be
  Prob\{  \sum_{i=1}^{n}S_{i}(0) . \theta_{i} ( \eta_{i}(t) + \frac{{\eta_{i}(t)}^{2}}{2} ) \leq  - VaR_{\alpha } \} = \alpha
\label{e7}
  ,\ee
 then an elementary mathematical tool give that :
 \be
\eta_{i}(t) + \frac{{\eta_{i}(t)}^{2}}{2} =\frac{1}{2}
{(\eta_{i}(t) +1)}^{2} - 1 \label{eq7} \ee
 therefore
\be
 Prob\{  \sum_{i=1}^{n} S_{i}(0) . \frac{\theta_{i}}{2}
{(\eta_{i}(t) +1)}^{2}     )   \leq     - VaR_{\alpha } +
\sum_{i=1}^{n} \frac{\theta_{i}}{2}\cdot S_{i}(0) \} = \alpha .
\label{e8}
 \ee
  By posing  $X =(\eta_{1}+1 , ... , \eta_{n}+1 ) $, it is straightforward that , X
is an elliptic distribution due to the fact  that it is a linear
combination of elliptic distribution $\eta $. We note $X\sim N(\mu
+1,\Sigma ,\phi^{'})$ with a continuous density function
$h_{1}(x)$.  Remark that $$\sum_{i=1}^{n} \alpha_{i} {(\eta_{i}(t)
+1)}^{2} =(x,\Lambda.x)$$ with  $\Lambda={( \alpha
_{ii})}_{i=1..n} $ is a diagonal matrix with diagonal values     $
\alpha_{ii}=\frac{ S_{i}(0) . \theta_{i}}{2} \geq 0$ and   $
\mu^{'} =(\mu_{1}+1,\ldots,\mu_{n} +1) =\mu  + \rm 1
\hspace{-0.25em} I$ is the mean vector of X , therefore (\ref{e8})
becomes
 \be Prob\{  (X,\Lambda.X) \leq      k
\}  =\alpha  \label{eq8} \ee

with $  k= -VaR_{\alpha }  + \sum_{i=1}^{n} \alpha_{ii}
=\frac{P(0)}{2}- VaR_{\alpha }      $.
     We will suppose that $k >0$, this means that the Value-at-Risk
     of our portfolio's is not greater than P(0)/2 .

     \begin{remark}
        We remark that to estimate the Value-at-Risk of a
        portfolio of securities (i.e equities), the computation of our model need as inputs the quantity $\theta_{i}$ and the initial security price
        $S_{i}(0)$ for each $i=1..n$ as given in (\ref{e1}). Recall
        that in literature,  the computation of Quadratic
        Delta-Gamma ($\Delta$-$\Gamma$) VaR need as inputs the sensitivity vector
        $\Delta$ and the sensitivity matrix $\Gamma$, because of
        the presence of derivatives products in the portfolio.
        \end{remark}

\section{Reduction to an Integral Equation}

In this section, we will reduce the problem of computation of the Value at Risk for quadratic portfolio
     of equities to the study of the asymptotic behavior of the density function distribution over the hyper-sphere.

\subsection{ Notions of Elliptic Distributions}

      The following  definitions will be given as in \cite{EMS}(2002) .
 \subsubsection{Spherical Distribution}

\begin{defi}

     A random vector $ X=(X_1 , X_2 ,..., X_n )^{t} $ has a spherical distribution if for every orthogonal map
     $U\in R^{n \times n}$      (i.e. maps satisfying   $  U U^t = U^t U = I_{n \times n}  ) $
 $$UX =_d   X .\footnote{ $=_d $  denote equality in
 distribution}$$
  we note that:  $ X\sim S_{n}(\phi ) $.If X has a density  f(x) then this is equivalent to $ f(x)=g(x^t x ) = g( \| x\| ^2 ) $  for some function
 $g: R_{+} \longrightarrow R_{+}$, so that the spherical distributions are best interpreted as those distributions
  whose density is constant on spheres.
\end{defi}

Elliptical distributions extend the multivariate  normal
$N_{n}(\mu ,\Sigma )$, for which $\mu$ is mean and $\Sigma $ is
the covariance matrix.  Mathematically, they are the affine maps
of spherical distributions in $\mathbb{R}^n$.

\subsubsection{Elliptic Distribution}
\begin{defi}
  Let   $ T: R^n \longrightarrow  R^n $ ,
             $   y   \longmapsto   Ay+\mu $ ,           $A \in R^{n\times n} $ , $ \mu \in R^{n} $  .
X  has an elliptical distribution if X = T ( Y ) and $Y\sim S_{n}
(\phi ) $.  If Y has a density $f(y) = g(y^t y) $ and if A is
regular (det(A) $\not = 0 $ so that $\Sigma=A^t A$ is strictly
positive),then $X= AY + \mu $ has a density
 \be
 h(x) = g((x-\mu )^t {\Sigma}^{-1}  (x-\mu )) /\sqrt{det(\Sigma)}     \label{pdf1}
\ee
 and the contours of equal density are now ellipsoids. An
elliptical distribution is fully described by its mean, its
covariance matrix and its characteristic generator.
\end{defi}

\begin{itemize}
\item  Any linear combination of an elliptically distributed
random vector is also elliptical with the same characteristic
generator $\phi$ . If   $Y\sim  N_{n}(\mu ,\Sigma ,\phi ) $  , $
b\in \mathbf{R}^m $  and $B \in \mathbf{R}^{m \times n} $  then
       $ B.Y + b \sim  N_{m}(B\mu + b ,B\Sigma B^{t} ,\phi ) $  .
\end{itemize}

\subsection{ \bf{ Integral Equation with solution VaR}     }

   Since X  is an elliptic distribution, its density take the following form:

$$ h_{1}(x) = h( (x-\rm 1 \hspace{-0.25em} I)) $$
$\rm 1 \hspace{-0.25em} I $ is the vector of unities and h is the density
function of $\eta $ which take the form :
  $$h(x) = g((x-\mu ) {\Sigma}^{-1}  (x-\mu )^{t})/\sqrt{det( \Sigma )} , $$
  therefore  we have the following equation
$$Prob\{  (X,\Lambda X ) \geq - VaR_{\alpha  } +\sum_{1}^n  \alpha_{ii}        \} = 1-\alpha= I(k) $$
with $VaR_{\alpha} $, as solution such that $ k=-VaR_{\alpha  }
+\sum_{1}^n \alpha_{ii} $.

In terms of our elliptic distribution parameters we have to solve
the following equation:

\be I(k)=  \int_{\{   (x,\Lambda .x ) \geq    k           \} }
h_{1}(x) dx =1-\alpha     \label{eq1} \ee

with $ X \sim E_{n}( \mu+1 , \Sigma , \phi ) $,  $ AA^t = \Sigma $
and  $I(k)$ given as follow:
$$I(k)=  \int_{\{ (x,\Lambda .x )   \geq   k   \} }  g((y-\mu -\rm 1 \hspace{-0.25em} I)^{t} {\Sigma}^{-1}  (y-\mu-\rm 1 \hspace{-0.25em} I ))     \frac{dy}{\sqrt{det( \Sigma )}}   $$
since $\Lambda $ is a diagonal matrix with all positive diagonal
values, we decompose $\Lambda = \Lambda^{1/2} \cdot \Lambda^{1/2}$
therefore  the equation (\ref{eq1}) becomes
 $$I(k)=  \int_{\{ <\Lambda^{1/2} (Az+\mu + \rm 1 \hspace{-0.25em} I),\Lambda^{1/2}  (Az + \mu + \rm 1 \hspace{-0.25em} I)>    \geq   k   \} }  g( \| z\|^{2}   )     dz       =  \int_{\ {\|   \Lambda^{1/2} (Az+\mu + \rm 1 \hspace{-0.25em} I)\|}_{2}^2     \leq  k   \} }  g( \| z\|^{2}   )     dz      . $$
 Cholesky decomposition states that  $ \Sigma = AA^{t} $ when $\Sigma$ is
suppose to be positive, therefore if we changing the variable
$z=A^{-1}(y-\mu -\rm 1 \hspace{-0.25em} I)$, the precedent
integral becomes :
$$I(k)=  \int_{\  (Az+\mu + \rm 1 \hspace{-0.25em} I)^{t} \Lambda (Az + \mu + \rm 1 \hspace{-0.25em} I)    \geq   k   \} }  g( \| z\|^{2}   )     dz     . $$
If we do the following decomposition $(Az+\mu + \rm 1
\hspace{-0.25em} I)^{t} \Lambda (Az + \mu + \rm 1 \hspace{-0.25em}
I)  =(z+v)^{t}D(z+v) + \delta  $, with  $D=A^{t}\cdot \Lambda
\cdot A$  ,   $ v=A^{-1}(\mu +\rm 1 \hspace{-0.25em} I ) $ and
$\delta =0$, after some elementary calculus
$$I(k)=  \int_{\{  (z+v)^{t}D(z+v)       \geq   k - \delta   \} }  g( \| z\|^{2}   )     dz      .$$
If we suggesting $k_{1}= k - \delta=R^2 $,  $ z+v=u $, $ dz = du$,
we find that
$$  I(k) = \int_{\{ u^{t}. D.u   \geq  R^2    \} }  g( \|u-v\|^{2} )    dz  = 1-\alpha  .      $$
By introducing the variable  $z=D^{1/2}u/R$, we have that:

$$ I(R) = R^n \int_{\{ \| z\|    \geq  1    \} }      g(\| R{D^\frac{-1}{2}}z-v\|^2 )  \frac{dz}{\sqrt{det(D)}}$$

next, by using spherical variables  $z=r.\xi $ with $\xi \in
S_{n-1}$ and  $ dz = r^{n-1} dr d\sigma(\xi )$, where  $d\sigma
(z) $ is a elementary surface of  z  on $ S_{n-1} = \{  \xi  | \xi
\in  {\mathbb{R}}^{n} , \xi _{1}^{2} + \xi_{2}^{2} + ... + \xi
_{n}^{2} = 1  \} $ and by introducing the function $J(r,R)$ such
that
 \be
 J(r, R) =  \int_{S_{n-1}}   g(\| rR{D^\frac{-1}{2}}\xi -v\|^2 )
d\sigma (\xi )  \label{eq10}
 \ee
 we obtain
\be
 R^{-n}\cdot I(R) =  \int_{1}^{\infty } r^{n-1} \Big{[} \int_{S_{n-1}}   g(\| rR{D^\frac{-1}{2}}\xi -v\|^2 )  d\sigma (\xi )  \Big{]} \frac{dr}{\sqrt{det(D)}}   =  \int_{1}^{\infty } r^{n-1} J(r,R) \frac{dr}{\sqrt{det(D)}}  . \label{eq0}
\ee

Next By  introducing the function
\be H(s)= s^{n} \int_{1}^{\infty
} r^{n-1} J(r,s) dr ,\label{funct} \ee

 our goal will be to solve the
following equation \be H(s)=(1-\alpha )\sqrt{det(D)}
.\label{equat}\ee

In the following section, we  propose to approximate J(r,R) by
applied the numerical methods giving in the paper of Alan Genz
(2003), ( see \cite{G} for more details).

\subsection{\bf{ Numerical approximation of J(r,R)} }

In this section, we estimate the integral J(r,R) by a numerical
methods given by  Alan Genz in \cite{G} .

\subsection{\bf{Some interpolation rules } on $S_{n-1}$}

   The paper \cite{G} of Alan Genz, give the following method.
  Suppose that we need to estimate the following integral
$$J(f)=\int_{S_{n-1}}  f(z) d\sigma (z) $$
where $d\sigma (z) $ is an element of surface on $S_{n-1} = \{ z |
z \in  {\mathbb{R}}^{n} , z_{1}^{2} + z_{2}^{2} + ... + z_{n}^{2}
= 1  \} $.

 In effect, let be the n-1 simplex by $T_{n-1}= \{  x |
x \in {\mathbb{R}}^{n-1} , 0 \leq x_{1} + x_{2} + ... + x_{n-1}
\leq 1 \} $ and for any  $  x\in  T_{n-1}$, define $x_{n} = 1-
\sum_{i=1}^{n-1} x_{i} $ . Also   $t_{p}=(
t_{p_{1}},...,t_{p_{n-1}})$   if points  $t_{0}$, $t_{1}$,\ldots ,
$t_{m}$    are given, satisfying the condition     :
$|t_{p}|=\sum_{i=1}^n  t_{p_{i}}  = 1 $  whenever   $\sum_{i=1}^n
p_{i} = m $, for non-negative integers  $p_{1}$,\ldots, $p_{n}$ ,
then the Lagrange interpolation formula (sylvester \cite{SY} for a
function g(x) on $T_{n-1}$ is given by

$$ L^{(m,n-1)} (g,x) = \sum_{|p|=m}  \prod_{i=1}^{n}   \prod_{j=0}^{p_{i}-1}  \frac{ x_{i}^{2} -  t_{j}^{2}}{ t_{p_{i}}^{2} -  t_{j}^{2}}    g(t_{p})  $$
$L^{(m,n-1)} (g,x)  $ is the unique polynomial of degree m which
interpolates g(x)  at all of the    $C_{ m+n-1}^m $  points in the
set  $\{x  | x=(t_{p_{1}},...,t_{p_{n-1}}) , |p|=m \}$. Silvester
provided families of points, satisfying the condition  $|t_{p}|=1
$ when $|p|=m$, in the form $ t_{i}=\frac{i+\mu}{m + \theta  n}$
for i=0,1,\ldots,m,  and $\mu$ real. If  $ 0\leq \theta \leq 1$ ,
all interpolation points for  $L^{(m,n-1)} (g,x) $ are in
$T_{n-1}$. Sylvester derived families of interpolators rules for
integration over $T_{n-1}$  by integrating  $L^{(m,n-1)} (g,x) $ .
    Fully symmetric interpolar integration rules can be obtained by  substitute  $  x_{i}=z_{i}^2 $ ,
     and $t_{i}=u_{i}^2 $ in $L^{(m,n-1)} (g,x) $, and define
$$M^{(m,n)} (f,z) =  \sum_{|p|=m}  \prod_{i=1}^{n}   \prod_{j=0}^{p_{i}-1}  \frac{ z_{i}^{2} - u_{j}^{2}}{ u_{p_{i}}^{2} -  u_{j}^{2}}    f\{u_{p}\}  $$
where $f\{u\}$  is a symmetric sum defined  by
$$f\{u\}= 2^{-c(u)} \sum_{s} f(s_{1}u_{1},s_{2}u_{2},\ldots,s_{n}u_{n})$$
with c(u) the number of nonzero entries in $(u_{1},\ldots,u_{n})$,
and the $\sum_{s} $ taken over all of the signs combinations that
occur when $s_{i}=\pm 1$ for those i with $u_{i}$ different to
zero.

\begin{lemma}
If
$$ w_{p} =J( \prod_{i=1}^{n}   \prod_{j=0}^{p_{i}-1}  \frac{ z_{i}^{2} - u_{j}^{2}}{ u_{p_{i}}^{2} -  u_{j}^{2}} )  $$
then
$$
 J(f) = R^{(m,n)}(f) =  \sum_{|p|=m}  w_{p} f\{u_{p}\}
$$

$$f\{u\}= 2^{-c(u)} \sum_{s} f(s_{1}u_{1},s_{2}u_{2},...,s_{n}u_{n})$$,
with c(u) the number of nonzero entries in $(u_{1},...,u_{n})$,
and the $\sum_{s} $ taken over all of the signs combinations that
occur when $s_{i}=\pm 1$ for those i with $u_{i}$ different to
zero.

\end{lemma}

The proof is given in \cite{G} by (Alan Genz (2003)) as follow:\\
  Let $z^{k}=z_{1}^{k_{1}}z_{2}^{k_{2}}\cdot z_{n}^{k_{n}}.$ J and R are both linear functionals, so it is sufficient to show that
$R^{(m,n)}(z^{k})= J(z^{k})$ whenever $|k|\leq 2m+1$.  If k has
any component $k_{i}$ that is odd, then  $J(z^{k})=0$, and
$R^{(m,n)}(z^{k})= 0$ because ever term $u_{q}^{k}$ in each of the
symmetry sums $f\{u_{p}\} $ has a cancelling term - $u_{q}^{k}$ .
Therefore, the only monomials that need to be considered are of
the form $z^{2k}$, with $|k|\leq m$. The uniqueness of
$L^{(m,n-1)} (g,x) $  implies $L^{(m,n-1)} (x^{k},x)=x^{k} $
whenever    $|k|\leq m$, so  $M^{(m,n)}(z^{2k},z)=z^{2k}$,
whenever  $|k|\leq m$.
Combining these results:\\
$$
 J(f)=M^{(m,n)}(f,z)) =\sum_{|p|=m} w_{p} f\{u_{p}\}
                                     =R^{(m,n)}(z^{k})
$$
whenever $f(z)=z^{k}$, with $|k|\leq 2m+1$, so  $R^{(m,n)}(f)$ has
polynomial degree 2m+1. For more details ( cf. Genz \cite{G}).

\subsection{Application to Numerical approximation of  $J(r,R)$}

    Since our goal is to estimate the integral (\ref{eq1}), it is straightforward that the theorem (4.2.1)
    is applicable to the function  f such that

 $$f(z) = g(\| rR{D^\frac{-1}{2}}z -v\|^2 ).$$

Next, since we have that
 $$f\{u_{p}\} = g(\| rR{D^\frac{-1}{2}}(s.u_{p})^{t} -v\|^2 ),$$
by introducing the approximate function $J_{u_{p}}$ that depend to
the choice of $ u_{p}$, (\ref{eq1}) becomes

\be J(r,R) \approx \sum_{|p|=m} \sum_{s} w_{p}  \quad  g(\|
rR{D^\frac{-1}{2}}(s.u_{p})^t -v\|^2 ) =J_{u_{p}}(r,R)
\label{approx}\ee

 we note $s.u_{p}={(s_{1}u_{1},...,s_{n}u_{n})}^t$.
\begin{remark}
 $ J_{u_{p}}(r,R)$ is the numerical approximation of $J(r,R)$ as
given in (\ref{approx}), is depend to the choice of interpolation
points $u_{p}$ on hypersphere. Recall that $J(r,R)$ was a fixed
function that depend to R and  the density function of our
elliptic distribution .
\end{remark}

By introducing $H_{u_{p}}$, the approximate
 function of $H$ as define in (\ref{funct}) that depend of  $J_{u_{p}}$,  such that

 \be H_{u_{p}}(s)=s^{n} \int_{1}^{\infty
} r^{n-1} J_{u_{p}}(r,s) dr \approx H(s) .\label{funct1} \ee

  By replace $H(s)$ in (\ref{equat}) by  $H_{u_{p}}(s)$ we then
prove the following result:
\begin{theorem}\label{Quad-VaR-elliptic}
   If we have a quadratic portfolio of securities (i.e equities)
   such that the Profit \& Loss function over the time
window of interest is, to good approximation, given by $\Delta \Pi
 \approx  \sum_{i=1}^{n} S_{i}(0)
\cdot\theta_{i} ( \eta_{i}(t) + \frac{{\eta_{i}(t)}^{2}}{2} ) $,
with portfolio weights $\theta_{i}$. Suppose moreover that the
joint log-returns is a random vector $(\eta_{1},\ldots,\eta_{n})$
that follows a continuous elliptic distribution, with probability
density as in $(\ref{pdf1})$,where $\mu$ is the vector mean and
$\Sigma$ is the variance-covariance matrix, and where we suppose
that $g(s^2 )$ is integrable over $\mathbb{R}$, continuous and
nowhere 0. Then the approximate portfolio's  quadratic elliptic
$VaR_{\alpha,u_{p}}^g$ at confidence ($1-\alpha$) is
   given by
   \be
    VaR_{\alpha,u_{p}}^g= \frac{P(0)}{2} - R_{g,u_{p}}^2
   \label{theo01}
   \ee
   where $R_{g,u_{p}}$ is the unique solution of the equation
\be   H_{u_{p}}(s) =(1-\alpha)\cdot \sqrt{det(D)}=
\frac{(1-\alpha)}{2^{n/2}}\sqrt{det(\Sigma)\prod_{i=1}^n
{\theta_{i}\cdot S_{i}}(0)}.\label{theo1} \ee

  In this case, we assume that our losses will not be greater than
half-price of the portfolio at time 0.
   \end{theorem}

   \begin{remark}
    The precedent theorem give to us an approximate Quadratic Portfolio Value-at-Risk ($VaR_{\alpha,u_{p}}^g$) that depend to our choice of interpolation
    points on hypersphere ,$\alpha$ and the function g. Therefore it is
    clear that the best choice of interpolation point will depend
    to the g function in (\ref{pdf1}).
    \end{remark}

 With some simple calculus we have
the following remark

\begin{remark}\label{rem}

\be J_{u_{p}}(r,R)= \sum_{|p|=m}  w_{p}   \sum_{s}
g\Big{(}a(s,u_{p},R)\cdot r^{2} -2\cdot b(s,u_{p},R,D,v)\cdot r +
c(v)\Big{)}
 \label{notJ} \ee
 for which
 $a(s,u_{p},R,D)=\|R{D^\frac{-1}{2}}(s.u_{p})\|^{2}$,
$b(s,u_{p},R,D,v)=R<{D^\frac{-1}{2}}(s.u_{p}),v>$,$ c=\|v\|^2$.
Sometimes, for more simplification we will note a,b,c.
 \end{remark}

   Since inequality of Schwartz give that $b^2-ac <0$ , we use the
   change of variable by posing $b1 =\frac{b^2 -ac}{a}<0$,
   $u=r-\frac{b}{a}$, by using   using the binom of Newton, and by introducing the function $G^{j,g}$ for $j=0,..,n-1$, such that we have the following
   remark
   \begin{remark}\label{rem1}

\be
  J(r,R)=\sum_{|p|=m}
w_{p} \sum_{s} \sum_{j=0}^{n-1} \binom{n-1}{j} {  (b/a)}^{n-1-j}
G_{u_{p},s}^{j,g} (R) \label{cour1}
 \ee
 with
 \be
G_{u_{p},s}^{j,g} (R)=\int_{1-\frac{b}{a}}^{\infty } z^j \cdot g(a
z^2 -b_{1} ) dz \label{note3}\ee
 for which a,b and c are defined in (\ref{rem}).
   \end{remark}

   By replace $d= b/a=\frac{<{D^\frac{-1}{2}}(s.u_{p}),v>}{R\|{D^\frac{-1}{2}}(s.u_{p})\|^{2}}$ by its value in
   (\ref{rem1}),we obtain
   \be
   G_{u_{p},s}^{j,g} (R)= \int_{1-\frac{<{D^\frac{-1}{2}}(s.u_{p}),v>}{R\|{D^\frac{-1}{2}}(s.u_{p})\|^{2}}}^{\infty }
z^j \cdot g \Big{(}R^2 \cdot\|{D^\frac{-1}{2}}(s.u_{p})\|^{2} z^2
-
\frac{<{D^\frac{-1}{2}}(s.u_{p}),v>^2}{\|{D^\frac{-1}{2}}(s.u_{p})\|^{2}}
+\|v\|^2  \Big{)} dz \label{not2} \ee

    we then have the following theorem
   \begin{theorem}
   If we have a quadratic portfolio of securities (i.e equities) for which the joint
   securities log-returns changes with continuous elliptic
   distribution with pdf distribution as in $(\ref{pdf1})$, then the approximate portfolio's  quadratic elliptic
$VaR_{\alpha,u_{p}}^g$ at confidence ($1-\alpha$) is
   given by
   \be
    VaR_{\alpha,u_{p}}^g= \frac{P(0)}{2} - R_{g,u_{p}}^2
   \label{theo01}
   \ee
   where $R_{g,u_{p}}$ is the unique solution of the equation
\be
 \sum_{|p|=m}  w_{p}   \sum_{s} \sum_{j=0}^{n-1}
\binom{n-1}{j} R^{j+1} \cdot
{(\frac{<{D^\frac{-1}{2}}(s.u_{p}),v>}{\|{D^\frac{-1}{2}}(s.u_{p})\|^{2}})}^{n-1-j}
\cdot G_{u_{p},s}^{j,g} (R) =
\frac{(1-\alpha)}{2^{n/2}}\sqrt{det(\Sigma)\prod_{i=1}^n
{\theta_{i}\cdot S_{i}}(0)} \label{cor1} .\ee
 In this case, we assume that our losses will not be greater than
half-price of the portfolio at time 0.
   \end{theorem}

   \begin{remark}
   We have reduced our problem to one dimensional integral equation. Therefore, to get an explicit equation to
   solve, we need to estimate $G_{u_{p},s}^{j,g} (R)  $ that depend to
   R with parameters $g$,$u_{p}$,v and D.
   \end{remark}

Therefore, in the case of normal distribution or $t$-distribution,
it will suffices to replace g in the expression of (\ref{note3}),
an to estimate the one dimensional integral (\ref{note3}).

\subsubsection{\bf{ The case of normal distribution}}

   In the case of normal distribution, the pdf is given by:
   \be f(x)=\frac{1}{\sqrt{(2\pi)^n |\Sigma|}}exp(-\frac{1}{2}
   (x-\mu)\Sigma^{-1}(x-\mu)^t )\label{pdfnorm}\ee
   and specific is given as follow
   $$ g(x)=(2\pi)^{-\frac{n}{2}} e^{-\frac{x}{2}}=C(n).e^{-\frac{x}{2}}$$
   therefore it suffices to replace g in ($\ref{note3}$) then

\be
 G_{u_{p},s}^j (R)=(2\pi)^{-\frac{n}{2}} e^{\frac{b_{1}}{2} }
\int_{1-\frac{b}{a}}^{\infty } u^j e^{-\frac{a u^2}{2}}  du
.\label{eq12} \ee

 If $1-\frac{b}{a} >0 $ ( it is the case when R is
 sufficiently big such that $|v| < R
 \|{D^\frac{-1}{2}}(s.u_{p})\|$ ).

\be
 \frac{G_{u_{p},s}^j (R)}{\exp({-\frac{\|v\|^2}{2}})(2\pi)^{-\frac{n}{2}}}=
 \exp{(\frac{<{D^\frac{-1}{2}}(s.u_{p}),v>^2}{R\|{D^\frac{-1}{2}}(s^{t}u_{p})\|^{2}})}
 {(2/a)}^{\frac{1+j}{2}} \Gamma(\frac{j+1}{2},\frac{(R\|{D^\frac{-1}{2}}(s^{t}u_{p})\|)^2}{2}{(1-\frac{<{D^\frac{-1}{2}}(s.u_{p}),v>}{R\|{D^\frac{-1}{2}}(s^{t}u_{p})\|^{2}})}^2
) \label{equa12} \ee

therefore, since $ a =(R\|{D^\frac{-1}{2}}(s.u_{p})\|)^2$ we have
the following theorem

\begin{theorem}
If we have a portfolio of securities (i.e equities), such that the
Profit \& Loss function over the time window of interest is, to
good approximation, given by $\Delta \Pi
 \approx  \sum_{i=1}^{n} S_{i}(0)
\cdot\theta_{i} ( \eta_{i}(t) + \frac{{\eta_{i}(t)}^{2}}{2} ) $,
with portfolio weights $\theta_{i}$. Suppose moreover that the
joint log-returns is a random vector $(\eta_{1},\ldots,\eta_{n})$
that follows a continuous multivariate normal distribution with
density function in (\ref{pdfnorm}), vector mean $\mu$ , and
variance-covariance matrix $\Sigma $, the Quadratic Value-at-Risk
$(VaR_{\alpha,u_{p}}$ at confidence $1 - \alpha $ is given by the
following formula
$$R_{u_{p},\alpha}^2 = -VaR_{\alpha,u_{p}}+\frac{P(0)}{2}$$
for which $R_{u_{p},\alpha}$ is the unique solution of the
following transcendental equation.

\be
2(1-\alpha)\frac{\sqrt{det(D)}}{(2\pi)^{\frac{n}{2}}}=\sum_{|p|=m}
w_{p} \sum_{s}\frac{(<{D^\frac{-1}{2}}(s.u_{p}),v>)^{(n-j-1)}
}{\|{D^\frac{-1}{2}}(s.u_{p})\|^{(2n-1-j)}}
e^{\frac{b_{1}}{2}}\sum_{j=0}^{n-1}  \binom{n-1}{j}
\Gamma(\frac{j+1}{2},\frac{a}{2}{(1-\frac{b}{a})}^2 )
\label{equa12} \ee
 for which
$b_{1}=\frac{(<{D^\frac{-1}{2}}(s.u_{p}),v>)^2 }
{\|{D^\frac{-1}{2}}(s.u_{p})\|^{2}}-\|v\|^2 $,
$\frac{b}{a}=\frac{<{D^\frac{-1}{2}}(s.u_{p}),v> }{R\|
D^\frac{-1}{2} (s.u_{p})\|^2 } $, $a=R^2
\|{D^\frac{-1}{2}}(s.u_{p})\|^{2}$. In this case, we implicitly
assume  $VaR_{\alpha;u_{p}} \leq P(0)/2$. $\Gamma$ is the
incomplete gamma function.
\end{theorem}

\subsubsection{\bf{Case of $t$-student distribution}}

If our elliptic distribution is in particular chosen as the
multivariate t-student distribution,  we will have density
function given by
\be g(x)=\frac{\Gamma (\frac{\nu +
n}{2})}{\Gamma(\nu/2). \pi^{n/2} }  {\Big{(}1+\frac{x }{\nu}
\Big{)}}^{(\frac{-\nu-n}{2})}=C(n,\nu){\Big{(}1+\frac{x }{\nu}
\Big{)}}^{(\frac{-\nu-n}{2})}\label{pdf3}
 \ee
 therefore by replacing g  in  ($\ref{cor1}$), we obtain the equation
\be
   \sum_{|p|=m}  w_{p}   \sum_{s} \sum_{j=0}^{n-1}
   \binom{n-1}{j} {(b/a)}^{n-1-j}
\int_{1-\frac{b}{a}}^{\infty } u^j {\Big{(}1+\frac{  a u^2 -
b_{1}^2 }{\nu} \Big{)}}^{(\frac{-\nu-n}{2})} du = \frac{(1-\alpha)
\sqrt{det(D)}}{ C(n,\nu)R^{n}}\label{equstud0} .\ee

suggesting $ c_{1}=\nu - b_{1}^2 $   , our equation is reduce to
\be
   \sum_{|p|=m}  w_{p}   \sum_{s} \sum_{j=0}^{n-1}
   \binom{n-1}{j} {(b/a)}^{n-1-j}
\int_{1-\frac{b}{a}}^{\infty } u^j {\Big{(}a u^2 +c_{1}
\Big{)}}^{(\frac{-\nu-n}{2})} du = \frac{(1-\alpha)
\sqrt{det(D)}}{\nu^{\frac{\nu+n}{2} }
C(n,\nu)R^{n}}\label{equstud1} .\ee

changing variable in this integral according to $v=u^2 $ and
$\beta=\frac{a}{c_{1}}$, we find that
\be
  R^{n} \sum_{|p|=m}  w_{p}   \sum_{s} \sum_{j=0}^{n-1}
   \binom{n-1}{j} {(b/a)}^{n-1-j} c_{1}^{\frac{-n-\nu}{2}}
\int_{{(1-\frac{b}{a})^2}}^{\infty } v^{\frac{j+1}{2}-1}
{\Big{(}\beta v + 1 \Big{)}}^{(\frac{-\nu-n}{2})} du =
\frac{(1-\alpha) \pi^{n/2}\Gamma(\nu/2)
\sqrt{det(D)}}{\nu^{\frac{\nu+n}{2} } \Gamma (\frac{\nu + n}{2})
}\label{equstud1} .\ee

 For the latter integral equation, we will
use the following formula from \cite{GR}:

\begin{lemma} (cf. \cite{GR}, formula 3.194(2)).
If $|arg (\frac{u}{\beta})|<\pi $, and $Re(\nu_{1})>Re(\mu)>0$ ,
then \be \int_{u}^{+\infty}
    x^{\mu-1} (1 + \beta x)^{-\nu_{1}}
    dx= \frac{u^{\mu-\nu_{1}}
     \beta^{-\nu_{1}}}{\nu_{1} -\mu }
    {_2F}_{1}(\nu_{1},\nu_{1}-\mu;\nu_{1}-\mu+1;-\frac{1}{\beta \cdot
    u}).
    \label{rem13}                  \ee
Here $_2 F _1 (\alpha ; \beta , \gamma ; w ) $ is the
hypergeometric function.
    \end{lemma}
    In our case,$\nu_{1}=\frac{\nu +n}{2}$, $u=(1-b/a)^2$, $\nu_{1}-\mu=\frac{n +\nu
-j-1}{2}$, $\nu_{1}-\mu+1=\frac{n +\nu -j+1}{2}$ therefore
  If we replace in $(\ref{equstud1})$, we will
obtain the following result.

 \begin{theorem}
If we have a portfolio of securities (i.e equities), such that the
Profit \& Loss function over the time window of interest is, to
good approximation, given by $\Delta \Pi
 \approx  \sum_{i=1}^{n} S_{i}(0)
\cdot\theta_{i} ( \eta_{i}(t) + \frac{{\eta_{i}(t)}^{2}}{2} ) $,
with portfolio weights $\theta_{i}$. Suppose moreover that the
joint log-returns is a random vector $(\eta_{1},\ldots,\eta_{n})$
that follows a continuous multivariate $t$-distribution with
density function given by (\ref{pdf3}), vector mean $\mu$ , and
variance-covariance matrix $\Sigma $, the Quadratic Value-at-Risk
$(VaR_{\alpha,u_{p}}$ at confidence $1 - \alpha $ is given by the
following formula
$$R_{u_{p},\alpha}^2 = -VaR_{\alpha,u_{p}}+\frac{P(0)}{2}$$
for which $R_{u_{p},\alpha}$ is the unique solution of the
following transcendental equation.
   \be \frac{R^{n}}{(1-\alpha)} \sum_{|p|=m}  w_{p}
\sum_{s}\sum_{j=0}^{n-1}\binom{n}{j}\frac{(b/a)^{n-1-j}}{ {(\nu
-b_{1}^2 )}^\frac{-n-\nu}{2}} \frac{ {_2 F}_1 \left[ \frac{n+\nu
}{2 } , \frac{n +\nu-j-1 }{2 } ; \frac{n+\nu-j+1 }{2 } ; \frac{
b_{1}^2
-\nu}{a(1-\frac{b}{a})^2}\right]}{(n+\nu-j-1)\sqrt{det(\Sigma
)\prod_{i=1}^n {\theta_{i}\cdot S_{i}}(0)}} =\frac{
(\pi/2)^{n/2}\Gamma(\nu/2) }{\nu^{\frac{\nu+n}{2} } \Gamma
(\frac{\nu + n}{2}) }
 \label{th}\ee
 for which
$b_{1}=\frac{(<{D^\frac{-1}{2}}(s.u_{p}),v>)^2 }
{\|{D^\frac{-1}{2}}(s.u_{p})\|^{2}}-\|v\|^2 $,
$\frac{b}{a}=\frac{<{D^\frac{-1}{2}}(s.u_{p}),v> }{R\|
D^\frac{-1}{2} (s.u_{p})\|^2 } $, $a=R^2
\|{D^\frac{-1}{2}}(s.u_{p})\|^{2}$ . In this case, we implicitly
assume that $VaR_{\alpha,u_{p}}\leq P(0)/2 $
\end{theorem}
\begin{remark}
\rm{ Note that, Hypergeometric $_2 F _1 $'s have been extensively
studies, and numerical software for their evaluation is available
in Maple and in Mathematica.}
\end{remark}

\section{ Quadratic VaR with mixture of elliptic Distributions}

  Mixture distributions can be used to model situations where the
data can be viewed as arising from two or more distinct classes of
populations; see also \cite{MX}. For example, in the context of
Risk Management, if we divide trading days into two sets, quiet
days and hectic days, a mixture model will be based on the fact
that returns are moderate on quiet days, but can be unusually
large or small on hectic days. Practical applications of mixture
models to compute VaR can be found in Zangari (1996), who uses a
mixture normal to incorporate fat tails in VaR estimation. In this
section, we sketch how to generalize the preceding section to the
situation where the joint log-returns follow a mixture of elliptic
distributions, that is, a convex linear combination of elliptic
distributions.

\medskip
\begin{defi} \rm{We say that $ (X_{1},\ldots,X_{n})$ has a joint distribution
that is the mixture  of $q $ elliptic distributions
$N(\mu_{j},\Sigma_{j},\phi_{j}) $\footnote{or
$N(\mu_{j},\Sigma_{j},g_{j})$ if we parameterize elliptical
distributions using $g_{j} $ instead of $\phi_{j} $}, with weights
$\{\beta_{j}\}$  (j=1,..,q ;  $\beta_{j} > 0$ ;  $\sum_{j=1}^q
\beta_{j} = 1$), if its cumulative distribution function can be
written as
$$
F_{X_{1},\ldots,X_{n}}(x_{1},\ldots,x_{n}) = \sum_{j=1}^q
\beta_{j} F_{j}(x_{1},\ldots,x_{n})
$$
with $F_{j}(x_{1},\ldots,x_{n}) $ the cdf of
$N(\mu_{j},\Sigma_{j},\phi_{j}) $. }

 \end{defi}

\begin{remark} \rm{
In practice, one would usually limit oneself to $q = 2 $, due to
estimation and identification problems; see \cite{MX}. }
\end{remark}

We will suppose that all our elliptic distributions
$N(\mu_{j},\Sigma_{j},\phi_{j}) $
admit a pdf :\\
\be f_{j}(x)= |\Sigma _j |^{-1/2 } g_{j}((x-\mu_{j} )
{\Sigma_{j}}^{-1}  (x-\mu_{j} )^t ) \label{med} \ee
 for which each  $g_{j}$
is continuous integrable function over $\mathbb{R}$, and that the
$g_{j}$ never vanish  jointly in a point of $\mathbb{R}^q $.
 The pdf of the mixture
will then simply be $\sum _{j = 1 } ^q \beta _j f_j (x) . $
\medskip

Let
$$\Sigma_{j} = A_{j} ^t \; A_{j} $$

So, following (\ref{eq0}),we introduce $J_{k}(r,R)$ such that
 \be
\alpha R^{-n} = \sum_{k=1}^{q} \int_{1}^{\infty } r^{n-1} \Big{[}
\int_{S_{n-1}} g_{k}(\| rR{D_{k}^\frac{-1}{2}}\xi -v_{k}\|^2 )
d\sigma (\xi ) \Big{]} \frac{dr}{\sqrt{det(D_{k})}}
=\sum_{k=1}^{q} \int_{1}^{\infty } r^{n-1} J_{k}(r,R) dr
\label{ellip1} \ee

Next
 following (\ref{note3}), we introduce the function
  \be G_{u_{p},s,k}^{j,g}
(R)=R^{n}\int_{1-\frac{b_{k}}{a_{k}}}^{\infty } z^j \cdot
g_{k}(a_{k} z^2 -b_{1k} ) dz \label{note3mix}\ee with
$a_{k}=\|R{D^\frac{-1}{2}}(s.u_{pk})\|^{2}$,
$b_{k}=R<{D_{k}^\frac{-1}{2}}(s.u_{pk}),v>$,$
c_{k}=\|v_{k}\|^2$,$b_{1k} =\frac{b_{k}^2 -a_{k}c_{k}}{a_{k}}$,
then we have the following corollary

\begin{theorem}
If we have a portfolio of securities (i.e equities)
   such that the Profit \& Loss function over the time
window of interest is, to good approximation, given by $\Delta \Pi
 \approx  \sum_{i=1}^{n} S_{i}(0)
\cdot\theta_{i} ( \eta_{i}(t) + \frac{{\eta_{i}(t)}^{2}}{2} ) $,
with portfolio weights $\theta_{i}$. Suppose moreover that the
joint log-returns is a random vector $(\eta_{1},\ldots,\eta_{n})$
  is a mixture of $q$ elliptic distributions, with
 density
$$
h(x)=\sum_{j=1}^q  \beta _j {|\Sigma_{j}|}^{-1/2 }
g_{j}((x-\mu_{j} )\Sigma_{j}^{-1}(x-\mu_{j} )^{t})
$$
where $\mu_{j}$ is the vector mean, and  $\Sigma_{j} $ the
variance-covariance matrix of the $j $-th component of the
mixture.  We suppose that each $g_j $ is  integrable function over
$\mathbb{R}$, and that the $g_j $ never vanish jointly in a point
of $\mathbb{R }^m $. Then the value-at-Risk, or {\em Quadratic
mixture-elliptic VaR}, at confidence $1 - \alpha $ is given as the
solution of the transcendental equation
 \be
   \sum_{k=1}^{q}\sum_{|p|=m}  w_{p}   \sum_{s} \sum_{j=0}^{n-1}
\binom{n-1}{j} {(b_{k}/a_{k})}^{n-1-j}
\frac{G_{u_{p},s,k}^{j,g}\Big{(}(\frac{P(0)}{2}-VaR_{\alpha,u_{p}}^{g})^{1/2}
\Big{)}}{ \sqrt{det(\Sigma_{k})\prod_{i=1}^n {\theta_{i}\cdot
S_{i}}(0)}} = \frac{(1-\alpha)}{2^{n/2}} \label{corm1} \ee
 for which   $G_{u_{p},s,k}^{j,g}$ is
defined in (\ref{note3mix}).  In this case, we assume that our
losses will not be greater than half-price of the portfolio at
time 0.
   \end{theorem}
   \begin{remark} \rm{
One might, in certain situations, try to model with a mixture of
elliptic distributions which all have the same variance-covariance
and the same mean, and obtain for example a mixture of different
tail behaviors by playing with the $g_j $'s.}
\end{remark}

The preceding can immediately be specialized to a mixture of
normal distributions: the details will be left to the reader.

\subsection{\bf{Application with mixture of Student $t$-Distributions}}
We will consider a mixture of q Student $t$-distributions such
that, the $k^{th}$ density function $i=1,..,q $ will be given by
\be g_{k}(x)=\frac{\Gamma (\frac{\nu_{k} +
n}{2})}{\Gamma(\nu_{k}/2). \pi^{n/2} } {\Big{(}1+\frac{x
}{\nu_{k}}
\Big{)}}^{(\frac{-\nu_{k}-n}{2})}=C(n,\nu_{k}){\Big{(}1+\frac{x
}{\nu_{k}} \Big{)}}^{(\frac{-\nu_{k}-n}{2})}\label{pdfm3}
 \ee
 and $\Sigma_{k}=A_{k}^t A_{k}$  therefore by replacing $g_{k}$ by g  in  (\ref{th}) and since integration is a linear operation,
  we obtain the following theorem
\begin{theorem}
If we have a portfolio of securities (i.e equities)
   such that the Profit \& Loss function over the time
window of interest is, to good approximation, given by $\Delta \Pi
 \approx  \sum_{i=1}^{n} S_{i}(0)
\cdot\theta_{i} ( \eta_{i}(t) + \frac{{\eta_{i}(t)}^{2}}{2} ) $,
with portfolio weights $\theta_{i}$. Suppose moreover that the
joint log-returns is a random vector $(\eta_{1},\ldots,\eta_{n})$
  is a mixture of $q$ t-distributions, with
 density
$$
h(x)=\sum_{j=1}^q  \beta _j {|\Sigma_{j}|}^{-1/2 }\frac{\Gamma
(\frac{\nu_{j} + n}{2})}{\Gamma(\nu_{j}/2). \pi^{n/2} }
\Big{(}1+\frac{(x-\mu_{j} )\Sigma_{j}^{-1}(x-\mu_{j})^t }{\nu_{j}}
\Big{)}^{-\frac{n+\nu_{j}}{2}}
$$
where $\mu_{j}$ is the vector mean, and  $\Sigma_{j} $ the
variance-covariance matrix of the $j $-th component of the
mixture.  We suppose that each $g_j $ is  integrable function over
$\mathbb{R}$, and that the $g_j $ never vanish jointly in a point
of $\mathbb{R }^m $. Then the value-at-Risk, or {\em Quadratic
mixture-student VaR}, at confidence $1 - \alpha $ is given  by :
$$R_{u_{p}}^{2} =-VaR_{\alpha}+ \frac{P(0)}{2} $$
for which $R_{u_{p}}$ is the unique
positive solution of the following equation:
   \be \sum_{k=1}^{q}
   \sum_{|p|=m}\frac{w_{p}\Gamma (\frac{\nu_{k} + n}{2})}{\Gamma(\nu_{k}/2).
 }
\sum_{s}\sum_{j=0}^{n-1}\binom{n-1}{j}\frac{
R^{n}(b_{k}/a_{k})^{n-1-j}}{ {((\nu_{k} -b_{1k}^2 )/\nu_{k}
)}^\frac{-n-\nu}{2}} \frac{ {_2 F}_1 \left[ \frac{n+\nu_{k} }{2 }
, \frac{n +\nu_{k}-j-1 }{2 } ; \frac{n+\nu_{k}-j+1 }{2 } ; \frac{
b_{1k}^2
-\nu}{a_{k}(1-\frac{b_{k}}{a_{k}})^2}\right]}{(n+\nu_{k}-j-1)
\sqrt{|\Sigma_{k}|\prod_{i=1}^n {\theta_{i}\cdot S_{i}}(0)}}
=\frac{(1-\alpha) }{(\pi/2)^{\frac{-n}{2}}} \label{th}\ee
 for which
$b_{1k}=\frac{(<{D_{k}^\frac{-1}{2}}(s.u_{pk}),v_{k}>)^2 }
{\|{D_{k}^\frac{-1}{2}}(s.u_{pk})\|^{2}}-\|v_{k}\|^2 $,
$\frac{b_{k}}{a_{k}}=\frac{<{D_{k}^\frac{-1}{2}}(s.u_{pk}),v_{k}>
}{R\| D_{k}^\frac{-1}{2} (s.u_{pk})\|^2 } $, $a_{k}=R^2
\|{D_{k}^\frac{-1}{2}}(s.u_{pk})\|^{2}$ and
$det(\Sigma_{k})=|\Sigma_{k}|$ . In this case, we implicitly
assume that our losses will not be greater than $P(0)/2$.
\end{theorem}

\section{\bf{Elliptic Quadratic Expected Shortfall for  portfolio of securities}}
\medskip

Expected shortfall is a sub-additive risk statistic that describes
how large losses are on average when they exceed the VaR level.
Expected shortfall will therefore give an indication of the size
of extreme losses when the VaR threshold is breached. We will
evaluate the expected shortfall for a quadratic portfolio of
securities under the hypothesis of elliptically distributed risk
factors. Mathematically, the expected shortfall associated with a
given VaR is defined as:
$$
\mbox{Expected Shortfall } = \mathbb{E } (-\Delta \Pi \vert
-\Delta \Pi > VaR ),
$$
see for example \cite{MX}. Assuming again a multivariate elliptic
pdf $f(x) = {|\Sigma|}^{-1} g((x-\mu )\Sigma^{-1}(x-\mu )^{t}) $,
the Expected Shortfall at confidence level $1 - \alpha $ is given
by
\begin{eqnarray*}
- ES_{\alpha } &=& \mathbb{E } ( \Delta \Pi \mid   \Delta \Pi\leq
-VaR_{\alpha } ) \\
&=& \frac{1 }{\alpha } \mathbb{E } \left( \Delta \Pi \cdot 1_{\{
\Delta
\Pi \leq -VaR_{\alpha } \} } \right) \\
&=& \frac{1}{\alpha }\int_{\{ (x,\Lambda.x) -P(0)/2 \leq
-VaR_{\alpha }\}}
((x,\Lambda.x)-P(0)/2)  \ h_{1}(x) \ dx  \\
&=& \frac{ {|\Sigma|}^{-1/2}}{\alpha}  \int_{\{ (x,\Lambda.x) \leq
-VaR_{\alpha }+ P(0)/2\}} ((x,\Lambda.x)-P(0)/2)  \ g((x-\mu-1
)\Sigma^{-1}(x-\mu-1 )^{t}) dx .
\end{eqnarray*}
Using the definition of $VaR_{\alpha}$ and by replace $\Delta \Pi
=(X,\Lambda.X)-\frac{P(0)}{2}$, with random vector X define in
section 2,

\be ES_{\alpha }=\frac{P(0)}{2} -\frac{ {|\Sigma|}^{-1/2}}{\alpha}
\int_{\{ (x,\Lambda.x) \leq -VaR_{\alpha }+ P(0)/2\}}
(x,\Lambda.x) \ g((x-\mu-1 )\Sigma^{-1}(x-\mu-1 )^{t}) dx
\label{qes1} \ee

Let $\Sigma = A^t \; A $, as before.Doing the same linear changes
of variables as in section 2 and  section 3, we arrive at:
\begin{eqnarray*}
 ES_{\alpha } &=& \frac{P(0)}{2} -\frac{
{R^{n+2}|D|}^{-1/2}}{\alpha}\int_{0}^{1} r^{n+1}  \Big{[}
\int_{S_{n-1}}   g(\| rR{D^\frac{-1}{2}}\xi -v\|^2 )  d\sigma (\xi
)  \Big{]} dr \\
 &=& \frac{P(0)}{2}-\frac{ {R^{n+2}|D|}^{-1/2}}{\alpha}\int_{0}^{1
} r^{n+1} J(r,R)dr\\
&\approx &  \frac{P(0)}{2}-\frac{
{R^{n+2}|D|}^{-1/2}}{\alpha}\int_{0}^{1 } r^{n+1}
J_{u_{p}}(r,R)dr\\
 &=& \frac{P(0)}{2}-\frac{
{R^{n+2}|D|}^{-1/2}}{\alpha}\sum_{|p|=m} \sum_{s} w_{p}  \quad
\int_{0}^{1 } r^{n+1}  g(\| rR{D^\frac{-1}{2}}(s.u_{p})^t -v\|^2
)\quad dr \quad, \label{qes1}
\end{eqnarray*}
By introducing the function $Q_{u_{p},s}^{g} $ such that
\be
 Q_{u_{p},s}^{g}(R) =R^{n+2} \int_{0}^{1 } r^{n+1}  g(\| rR{D^\frac{-1}{2}}(s.u_{p})^t -v\|^2
)\quad dr \label{esquad2} \ee
 we have the following theorem

\begin{theorem} \label{EllipticES}
Suppose that the portfolio is quadratic in the risk-factors $X =
(X_1 , \cdots , X_n )$: $\Delta \Pi=(X,\Lambda \cdot
X)-\frac{P(0)}{2} $ and that $X \sim N(\mu+1 ,\Sigma ,\phi ) $,
with pdf $f(x)= {|\Sigma|}^{-1} g((x-\mu-1 )\Sigma^{-1}(x-\mu-1
)^{t}) $. If the $VaR_{\alpha}$ is given, then the expected
Shortfall at level $\alpha $ is given by :\\
\be ES_{\alpha }
=\frac{P(0)}{2}-\frac{{|D|}^{-1/2}}{\alpha}\sum_{|p|=m} \sum_{s}
w_{p} \quad Q_{u_{p},s}^{g}\Big{(}(\frac{P(0)}{2}-VaR_{\alpha})
^{1/2}\Big{)} . \label{EESformula} \ee
\end{theorem}
we introduce  $I_{1}^g $ and $I_{2}^g$ such that
$$ R^{-n-2}Q_{u_{p},s}^{g}(R)=\int_{0}^1 r^{n+1} g(a r^2 -2 b r +c )dr =\int_{0}^\infty -\int_{1}^\infty =I_{1,u_{p},s}^g (R) - I_{2,u_{p},s}^g (R) $$
Following the Integral (\ref{note3})

$$I_{2,u_{p},s}^g (R)=\sum_{j=0}^{n+1} \binom{n+1}{j} (b/a)^{n+1-j}\cdot G_{u_{p},s}^{j,g} (R)$$
 for which a,b and c are defined in remark (\ref{rem}) and
$$I_{1,u_{p},s}^g (R)=\int_{0}^\infty  r^{n+1} g(a r^2 -2 b r +c)dr$$

\subsection{Expected Shortfall with normal distribution}
 In the case of normal distribution, the pdf is given by
 (\ref{pdfnorm}) and the specific g is given as follow
   $$ g(x)=(2\pi)^{-\frac{n}{2}} e^{-\frac{x}{2}}=C(n).e^{-\frac{x}{2}}$$
   therefore it suffices to replace g in ($\ref{note3}$) then

\begin{eqnarray*}
 G_{u_{p},s}^j (R)&=&(2\pi)^{-\frac{n}{2}} \exp{\Big{(}\frac{b^2 -a c}{2a}\Big{) }}
\int_{1-\frac{b}{a}}^{\infty } u^j \exp{(-\frac{a u^2}{2})} du\\
&=&(2\pi)^{-\frac{n}{2}} \exp{\Big{(}\frac{b^2 -a c}{2a}\Big{) }}
{(2/a)}^{\frac{1+j}{2}}
\Gamma\Big{(}\frac{j+1}{2},\frac{a}{2}(1-\frac{b}{a})^2 \Big{)}
.\label{eqes12}
\end{eqnarray*}
By using the following lemma
\begin{lemma} (cf. \cite{GR}, formula 3.462(1)).
If  $Re(\nu) >0 $, and  $Re(\beta)>0$ , then \be
\int_{0}^{+\infty}
    x^{\nu-1} \exp{(-\beta x^2 - \lambda x)} dx = {(2\beta)}^{-\nu/2}
     \Gamma{(\nu)} \exp{\Big{(}\frac{\lambda^2}{8\beta}\Big{)}}
     \mathbb{D}_{-\nu}\Big{(}\frac{\lambda}{\sqrt{2\beta}}\Big{)}
    \label{rem3}                  \ee
Here $\mathbb{D}_{-\nu} $ is the parabolic cylinder function with
$$
\mathbb{D}_{-\nu}(z)=2^{\frac{-\nu}{2}} e^{\frac{-z^2}{2}}
\Big{[}\frac{\sqrt{\pi}}{\Gamma(\frac{1+\nu}{2})}\Phi\Big{(}\frac{\nu}{2},\frac{1}{2};\frac{
z^2}{2}\big{)}-\frac{\sqrt{2\pi}z}{\Gamma(\frac{\nu}{2})}\Phi
\Big{(}\frac{1+\nu}{2},\frac{3}{2};\frac{ z^2}{2}\big{)}\Big{]}
$$
where $\Phi$ is the confluent hypergeometric function (for more
details see \cite{GR} page 1018).
    \end{lemma}
 we next obtain
$$I_{2,u_{p},s}^g (R)=(2\pi)^{-\frac{n}{2}}\sum_{j=0}^{n+1} \binom{n+1}{j} (b/a)^{n+1-j}\cdot \exp{\Big{(}\frac{b^2 -a c}{4a}\Big{) }} {(2/a)}^{\frac{1+j}{2}}
\Gamma\Big{(}\frac{j+1}{2},\frac{a}{2}(1-\frac{b}{a})^2 \Big{)}$$
and
\begin{eqnarray*}
 I_{1,u_{p},s}^g (R)&=& (2\pi)^{-\frac{n}{2}} \exp(-\frac{\|v\|^2}{2}) \int_{0}^1
r^{n+1} \exp{\Big{(}-\frac{a r^2 -2 b r}{2}\Big{)}} dr\\
&=& (2\pi)^{-\frac{n}{2}}\exp(-\frac{\|v\|^2}{2})
{a}^{\frac{n+2}{2}}\Gamma(n+2) \exp{\Big{(}\frac{b^2}{4 a}\Big{)}}
\mathbb{D}_{-n-2}\Big{(}\frac{-b}{\sqrt{a}}\Big{)}
\end{eqnarray*}
for which $\mathbb{D}_{-n-2}$ is the parabolic cylinder function.
We have
therefore prove the following result:\\
\begin{theorem}
Suppose that the portfolio is quadratic in the risk-factors
$X=(X_1 , \cdots , X_n )$: \\
$\Delta \Pi=(X,\Lambda \cdot
X)-\frac{P(0)}{2} $ and that $X$ is a multivariate normal
distribution, If the $VaR_{\alpha}$ is given, then the expected
Shortfall at level $\alpha $ is given by :\\
\be ES_{\alpha }
=\frac{P(0)}{2}-\frac{R^{n+2}{|D|}^{-1/2}}{\alpha}\sum_{|p|=m}
\sum_{s} w_{p}\Big{[}I_{1,u_{p},s}^g (R)-I_{2,u_{p},s}^g
(R)\Big{]} . \label{EEeSformula} \ee
for which
$R=\sqrt{\frac{P(0)}{2}-VaR_{\alpha}}$
\end{theorem}

The preceding can immediately be specialized to a mixture of
normal distributions. The details will be left to the reader.

\subsection{Student $t$-distribution Quadratic Expected Shortfall}
Following the precedent section 3,4,5, and particularly the lemma
(\ref{rem13}) , the application can be specialized to a Student
$t$-distribution. The details will be left to the reader.

   \subsection{\bf{ How to choose an interpolation points $u_{p}$ on  hypersphere}}
    In order to obtain a good approximation of our integral, one
    will choose the points of interpolation $u_{p}$ of our g
    function such that our approximation is the best as possible.
    In the case where the g function  decreases quickly with all its derivatives of all order, in inspiring
    of the classical analysis, one will choose the points which
    render the maximum function $\|r R D^{-1/2}(s.u_p )- v\|$.
\section{\bf{ Conclusion}}
    By following the notion of Delta-Gamma Portfolio that contains derivatives instruments, we have introduced
    a Quadratic Portfolios  of securities (i.e equities) without the use of Delta and Gamma.
    By using the assumption that the joint securities log-returns follow a mixture of elliptic distributions,
    we have reduced the estimation of  VaR of such quadratic portfolio, to the resolution of a
    multiple integral equation, that contain a multiple integral over hypersphere.
    To approximate a multiple integral over hypersphere, we propose to use a numerical approximation method given by
    Alan Genz in\cite{G}. Therefore, the estimation of VaR is reduced to the resolution of
       one dimensional integral equation. To illustrate our method, we  give special attention to mixture of normal
distribution and mixture of  multivariate t-student distribution.
In the case of t-distribution, we need the hypergeometric special
function. For given VaR, we also show how to estimate the expected
shortfall of the Quadratic portfolio without derivatives
instruments, when the risk factors  follow an elliptic
distributions and we illustrate our proposition with normal
distribution by using the parabolic cylinder function. Note that
this method will be applicable to capital allocation, if we could
consider an institution as a portfolio of multi-lines businesses.
In the sequel paper, we will dealt with this numerical Quadratic
method, when the Portfolio contains derivatives instruments (i.e
options).  A concrete application need the estimation of $w_{p}$
such that $|p|=m$, therefore we send the reader to Alan Genz
\cite{G}.
       \appendix

\end{document}